\documentclass[aps,pra,twocolumn,showpacs,superscriptaddress]{revtex4}
\usepackage{graphicx}
\usepackage{amsmath}
\usepackage{bm}
\usepackage{float}
\usepackage{color}

\begin{document}
\title{A binary mixture of spinor atomic Bose-Einstein condensates}
\author{Z. F. Xu}
\affiliation{Center for Advanced Study, Tsinghua University, Beijing 100084,
People's Republic of China}
\author{Yunbo Zhang}
\affiliation{Institute of Theoretical Physics, Shanxi University,
Taiyuan 030006, People's Republic of China}
\author{L. You}
\affiliation{School of Physics, Georgia Institute of Technology,
Atlanta, Georgia 30332, USA}
\affiliation{Center for Advanced Study, Tsinghua University, Beijing 100084,
People's Republic of China}

\date{\today}

\begin{abstract}
We study the ground state and classify its phase diagram for a
mixture of two spin-1 condensates in the absence of external
magnetic (B-) field according to atomic parameters for intra- and
inter-species spin exchange coupling and singlet pairing
interaction. Ignoring the inter-species singlet pairing interaction,
the ground state phases are found analytically. Numerical approach
of simulated annealing is adopted when the singlet pairing
interaction is present. Our results on the phase diagram and the
boundaries between phases allow for easy identifications of quantum
phase transitions, that can be induced through the tuning of optical
traps and atom numbers. They provide the first insight and guidance
for several ongoing experiments on mixtures of spinor condensates.
\end{abstract}


\maketitle

The liberation of atomic hyperfine spin from optical trapping
opens the study of spin-dependent phenomena in atomic quantum gases
 \cite{kurn98,stenger98,barret01}. In the simplest case
of a spin-1 condensate, two quantum phases: polar and ferromagnetic,
exist depending on the sign of atomic spin exchange
interaction \cite{ho98,ohmi98,law98}.
The spin-2 system accompanied by richer physics is significantly
more complicated as in other higher spin systems \cite{ciobanu00,koashi00,ueda02,barnett06}.
Recently several groups have initiated the study of
spin-3 systems, of which the atomic $^{52}$Cr is a viable
experimental candidate \cite{santos06,diener06}.
Parallel to the efforts on spinor condensates are active
studies of condensate mixtures with more than one atomic species
or state \cite{ho96,myatt97,hall98,pu98,timmermans98,esry98,modugno02}.
Earlier works on double condensates rely on atoms with two
almost identically trapped internal states \cite{myatt97,hall98},
while more recent experiments demonstrated mixed condensates with tunable
inter-species interactions \cite{thalhammer08,papp08}.
Although the field is blossom with extensive studies on spinor condensates
and mixtures of scalar condensates, few have touched the subject of
 mixtures of spinor condensates \cite{luo07}.

This work concerns a binary mixture of spin-1 condensates,
with the mixture of $^{23}$Na and $^{87}$Rb atoms being a special case \cite{luo07}.
We study and identify the ground state phase diagram
under the mean-field approximation and the single spatial mode approximation (SMA)
for each of the two spinor condensates.
Our theory can be extended to more general mixtures
of higher spin condensates or mixtures with more than two constituents.

The interaction between two
different spin-1 atoms is described by the contact
pseudo-potential
${V}_{12}(\vec{r}_1-\vec{r}_2)=({g}_0^{(12)}\mathcal{P}_0
+{g}_1^{(12)}\mathcal{P}_1+{g}_2^{(12)}\mathcal{P}_2)\delta(\vec{r}_1-\vec{r}_2)$,
where ${g}_{0,1,2}^{(12)}=4\pi\hbar^2{a}_{0,1,2}^{(12)}/\mu$,
${a}_{0,1,2}^{(12)}$ are s-wave scattering lengthes
in the channel of total spin $F_{\rm tot}=0,1,2$
respectively. $\mu$ is the reduced mass.
$\mathcal{P}_{0,1,2}$ is the corresponding projection operator.
Using $\vec{F}_1\cdot\vec{F}_2=\mathcal{P}_2-\mathcal{P}_1-2\mathcal{P}_0$ \cite{ho98,ohmi98},
we find
\begin{eqnarray}
{V}_{12}(\vec{r}_1-\vec{r}_2)=(\alpha+\beta\mathbf{F}_1\cdot\mathbf{F}_2+\gamma\mathcal{P}_0)
  \delta(\vec{r}_1-\vec{r}_2),
  \label{intmsc}
\end{eqnarray}
where $\alpha=({g}_1^{(12)}+{g}_2^{(12)})/2$,
$\beta=(-{g}_1^{(12)}+{g}_2^{(12)})/2$, and
$\gamma=(2{g}_0^{(12)}-3{g}_1^{(12)}+{g}_2^{(12)})/2$.
Between same species spin-1 atoms, the interaction takes the familiar form
$V_{1,2}(\vec{r}_1-\vec{r}_2)=(\alpha_{1,2}+\beta_{1,2}\mathbf{F}_{1,2}\cdot\mathbf{F}_{1,2})
\delta(\vec{r}_1-\vec{r}_2)$, which respects the identical particle
symmetry, thus does not include the odd symmetry term of total spin 1
projection \cite{ho98,ohmi98}.

When atom numbers for both condensates are large, mean-field
approximations can be satisfactorily applied. We further assume
atomic interaction parameters are such that
SMA \cite{law98,yi02} for each spinor condensate
holds, {\it i.e.}, the mean-field mode functions are almost identical for
the three components of each spinor condensate.
We thus take
$\hat{\Psi}_i(\vec{r})=\sqrt{N_1}\,\psi(\vec{r})\zeta_i^{(1)}$ and
$\hat{\Phi}_i(\vec{r})=\sqrt{N_2}\,\phi(\vec{r})\zeta_i^{(2)}$,
respectively, for the mean fields of the two condensates. The mode
functions $\psi(\vec{r})$ and $\phi(\vec{r})$,
determined presumably by the stronger density dependent interactions,
are normalized according to $\int
d\vec{r}|\psi(\vec{r})|^2=\int d\vec{r}|\phi(\vec{r})|^2=1$.
$N_1$ and $N_2$ denote total atom numbers for the two condensates,
distributed into the respective spinor component with $N_i^{(1,2)}$
($\sum_i N_i^{(1,2)}=N_{1,2}$).
The effective order parameters for the spinor fields are
$\zeta^{(1)}_i=\sqrt{N_i^{(1)}/N_1}\, e^{-i\theta_i}$ and
$\zeta^{(2)}_i=\sqrt{N_i^{(2)}/N_2}\, e^{-i\varphi_i}$
with $\theta_i$ and $\varphi_i$ real phase values.
Neglecting spin-independent terms, the
Hamiltonian for our model of a mixture of two spin-1 condensates
takes the form
\begin{eqnarray}
  H&=&\frac{1}{2}{N_1^2C_1\beta_1}\mathbf{f}_1^2
  +\frac{1}{2}{N_2^2C_2\beta_2}\mathbf{f}_2^2\nonumber\\
  &&
  +\frac{1}{2}N_1N_2C_{12}\beta\mathbf{f}_1\cdot\mathbf{f}_2
  +\frac{1}{6}N_1N_2C_{12}\gamma|s_-|^2. \hskip 24pt
  \label{siener}
\end{eqnarray}
The interaction coefficients are
$C_1=\int d\vec{r}|\psi(\vec{r})|^4$, $C_2=\int
d\vec{r}|\phi(\vec{r})|^4$, and $C_{12}=\int
d\vec{r}|\psi(\vec{r})|^2|\phi(\vec{r})|^2$.
$f_{1i}=\sum_{m,n}\zeta^{(1)*}_mF^i_{mn}\zeta^{(1)}_n$,
$f_{2i}=\sum_{m,n}\zeta^{(2)*}_mF^i_{mn}\zeta^{(2)}_n$, and
$s_-=\sum_m(-1)^m\zeta^{(1)}_m\zeta^{(2)}_{-m}$.
Whether the two condensates are miscible, {\it i.e.}, completely
overlap or not, is unimportant. As long as the SMA holds
for each spinor condensate, our model retains its validity
even for immiscible condensates with diminishing cross spin coupling
$\propto C_{12}$.

Given the interaction coefficients, the above Hamiltonian (\ref{siener})
describes a system of two interacting spins.
It involves a total of six complex variables
when treated semi-classically.
The U(1) gauge symmetries for the two condensates give rise to two
constraints: the conservation of total atom numbers in each
condensate. After a detailed calculation, three
independent relative phases appear which we chose as:
 $\eta_1=\theta_1+\theta_{-1}-2\theta_0$,
$\eta_2=\varphi_1+\varphi_{-1}-2\varphi_0$, and
$\eta_3=\theta_{-1}+\varphi_1-\varphi_{-1}-\theta_{1}$.

To find the global ground state regardless of magnetization,
we minimize Eq. (\ref{siener})
with respect to $N^{(1)}_i$, $N^{(2)}_i$, and the
relative phases $\eta_{1,2,3}$ defined above.
Alternatively, we can minimize with respect to
semi-classical spin variables.
Both approaches give the same results,
provided that $\zeta_i^{(1,2)}$ is replaced by
the normalized value $\zeta^{(1,2)}_i/\sqrt{\sum_i N^{(1,2)}_i/N_{1,2}}$
to take into account the constraints explicitly.
The multi-dimensional minimization is numerically carried out with
the method of simulated annealing \cite{kirkpatrick83} where
analytical approach fails.
Without loss of generality,
we report the special case of $N_1=N_2=N$ in the following.

\begin{figure}[H]
\centering
\includegraphics[width=2.4in]{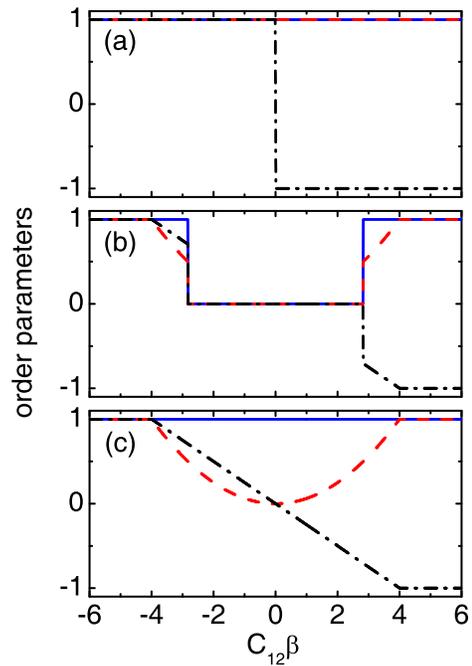}
\caption{(Color online). The dependence of ground state order
parameters on
$C_{12}\beta$ at fixed values of $C_1\beta_1$ and $C_2\beta_2$, for
$\gamma=0$. Blue solid lines, red dashed lines, and black dot-dashed
lines denote respectively the mean field order parameters $\mathbf{f}_1^2$,
$\mathbf{f}_2^2$, and $\mathbf{f}_1\cdot\mathbf{f}_2$. The three
subplots refer to the three cases of fixed intra-species spin exchange
interactions ($C_1\beta_1$, $C_2\beta_2$)$=$: (a) ($-1$,$-2$); (b)
($1,2$); and (c) ($-1,2$). } \label{fig1}
\end{figure}

We first discuss the simple situation of $\gamma=0$. The
results will serve as reference states for the complete
ground state phase diagram. As the ground
state is degenerate, it is convenient to restrict
into the subspace of $f_{i+}=f_{i-}=0\ (i=1,2)$, which leaves
two unknowns: $f_{1z}$ and $f_{2z}$. The ground state is
analytically derived including explicit formulae
for phase boundaries as functions of spin coupling
parameters. Figure \ref{fig1} illustrates the dependence of
order parameters on the inter-species
interaction $C_{12}\beta$ at fixed values of
intra-species spin exchange interaction parameter
$C_1\beta_1$ and $C_2\beta_2$, which
categorize the ground state into three cases.

First, when the two
spin-1 condensates are both ferromagnetic, the ground state is
simple, as shown in Fig. \ref{fig1}(a).  Both spins
are polarized: $\mathbf{f}_1^2=\mathbf{f}_2^2=1$. When
$C_{12}\beta>0$, they are anti-aligned as
$\mathbf{f}_1\cdot\mathbf{f}_2=-1$.
When $C_{12}\beta<0$, they are aligned with $\mathbf{f}_1\cdot\mathbf{f}_2=1$.

Secondly when both condensates are polar with
$C_1\beta_1>0$ and $C_2\beta_2>0$, we
assume $C_1\beta_1\neq C_2\beta_2$ without loss of generality.
The ground state is found to exhibit
several phases. If $C_{12}|\beta|<2\sqrt{C_1\beta_1C_2\beta_2}$,
{\it i.e.}, for weak intra-species spin exchange interaction, the two spins
become essentially independent, giving rise to
 $\mathbf{f}_1^2=0$, $\mathbf{f}_2^2=0$, and $\mathbf{f}_1\cdot\mathbf{f}_2=0$.
With increasing strength of $C_{12}|\beta|$, one of the spins
become polarized and the other is partially
polarized for $2\sqrt{C_1\beta_1C_2\beta_2}<C_{12}|\beta|<2
\max(C_1\beta_1, C_2\beta_2)$. The exact value for the resulting partial
polarization is determined by the relative strengths of the two
intra-species spin-exchange interactions. For
$C_1\beta_1<C_2\beta_2$, we have $\mathbf{f}_1^2=1$,
$\mathbf{f}_2^2=\big(C_{12}\beta/2C_2\beta_2\big)^2$, and
$\mathbf{f}_1\cdot\mathbf{f}_2=-C_{12}\beta/2C_2\beta_2$.
At stronger inter-species spin
exchange interaction, when $C_{12}|\beta|>2\max(C_1\beta_1,
C_2\beta_2)$, the spin-spin coupling between the two condensates
causes all atoms to be polarized as illustrated in Fig.
\ref{fig1}(b).

The third case as shown in Fig. \ref{fig1}(c), with one
ferromagnetic ($C_1\beta_1<0$) and one polar condensate ($C_2\beta_2>0$),
is found to be similar to case (b) of two polar condensates. When
$C_{12}|\beta|<2\max(C_1\beta_1, C_2\beta_2)$, the ground state
gives $\mathbf{f}_1^2=1$, $\mathbf{f}_2^2=\big(C_{12}\beta/2C_2\beta_2\big)^2$, and
$\mathbf{f}_1\cdot\mathbf{f}_2=-C_{12}\beta/2C_2\beta_2$,
which is the same as in case (b). When
$C_{12}|\beta|>2\max(C_1\beta_1, C_2\beta_2)$, we find
$\mathbf{f}_1^2=1$, $\mathbf{f}_2^2=1$, and
$|\mathbf{f}_1\cdot\mathbf{f}_2|=1$. We note that
Luo {\it et. al.} \cite{luo07}, recently performed quantum
calculations and studied the energy band
structure of essentially the same model for this case using
estimated inter-species atomic parameters for $^{87}$Rb and $^{23}$Na
atoms \cite{weiss03,pashov05}.

\begin{figure}[H]
\centering
\includegraphics[width=3.0in]{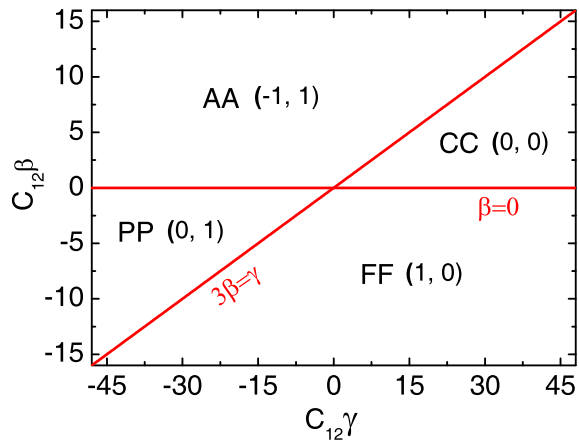}
\caption{(Color online). The ground state phase diagram for our
model when only inter-species spin exchange and singlet pair pairing
interactions are present. Red solid lines denote discontinuous phase
transition boundaries. The PP, FF, CC, and AA phases are denoted by
the values for the two order parameters
($\mathbf{f}_1\cdot\mathbf{f}_2$,$|s_-|$)=: (0,1), (1,0), (0,0),
and (-1,1) respectively.} \label{fig2}
\end{figure}

Next, we discuss the general case of $\gamma\neq 0$. Before
considering the full Hamiltonian (\ref{siener}), we illustrate the phase
diagram for the special case of $C_1\beta_1=C_2\beta_2=0$ in Fig.
\ref{fig2}. Only two terms remain in the Hamiltonian, which
formally resembles the Hamiltonian of a spin-2 condensate \cite{ciobanu00}.
The phases PP, FF, CC, and AA are characterized by the two order parameters
($\mathbf{f}_1\cdot\mathbf{f}_2=0$, $|s_-|=1$),
($\mathbf{f}_1\cdot\mathbf{f}_2=1$, $|s_-|=1$),
($\mathbf{f}_1\cdot\mathbf{f}_2=0$, $|s_-|=0$), and
($\mathbf{f}_1\cdot\mathbf{f}_2=-1$, $|s_-|=1$), respectively. The
other two order parameters take fixed values
($\mathbf{f}_1^2=0$,
$\mathbf{f}_2^2=0$), ($\mathbf{f}_1^2=1$, $\mathbf{f}_2^2=1$), and
($\mathbf{f}_1^2=1$, $\mathbf{f}_2^2=1$) respectively in the PP, FF, and AA
phases, while for the CC phase, their values remain undetermined.
Compared to a spin-2 condensate, we find
the PP, FF, and CC phases are similar to the polar, ferromagnetic,
and cyclic phases respectively.
There exists no counterpart for the AA phase
in a spin-2 condensate. The two hypothetical (spin-1) sub-spins
$\mathbf{f}_1$ and $\mathbf{f}_2$ making up
the total spin-2 are constrained to be identical.
In the case of a mixture as we discuss here,
however, the two spins are different.
The phase boundaries are specified by two lines
with $\beta=0$ and $3\beta=\gamma$ as in a spin-2
condensate. The transitions between different phases are all
first order in this case.

\begin{figure}[H]
\centering
\includegraphics[width=3.0in]{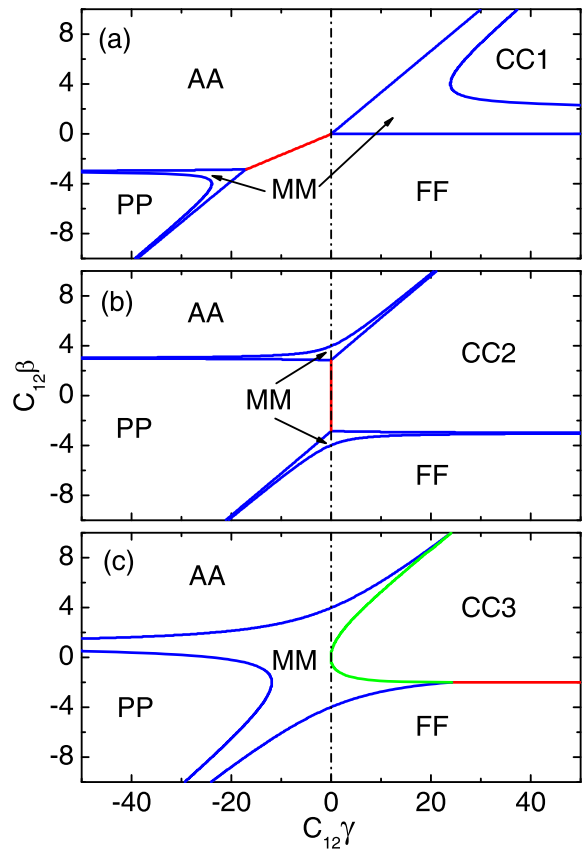}
\caption{(Color online). The ground state phase diagram of
our model at fixed values of
$C_1\beta_1$, $C_2\beta_2$. Blue solid lines denote continuous phase
transition boundaries. Red solid lines denote discontinuous phase
transition boundaries between two phases with fully determined order
parameters. The green solid line denotes the discontinuous phase
transition boundary between the fully determined phase CC3 and the
MM phase. The black dash-dotted lines correspond to $C_{12}\gamma=0$,
which serve as guides for the eye. The three subplots denote
fixed intra-specie spin exchange interaction
parameters as in Fig. \ref{fig1},
($C_1\beta_1$, $C_2\beta_2$)$=$: (a) ($-1$,$-2$); (b)
($1,2$); and (c) ($-1,2$). } \label{fig3}
\end{figure}

The most general case of our model Hamiltonian (\ref{siener})
is when all terms are present.
As before for $\gamma=0$, we will restrict to respective cases
specified by the fixed values for
$C_1\beta_1$ and $C_2\beta_2$.
Using full numerical simulations, we find the dependence
of ground state phases on
the four parameters: $C_1\beta_1$, $C_2\beta_2$, $C_{12}\beta$,
and $C_{12}\gamma$. The competitions among the first three
determine the ground state as categorized in Fig. \ref{fig1}.
The $\gamma$-term encourages pairing two different type of atoms
into singlets when $\gamma<0$; it minimizes $|s_-|$ instead when $\gamma>0$.
If $C_{12}|\gamma|$ is small enough compared to the first three
parameters, then intuitively we expect the ground state structure of
Fig. \ref{fig1} will largely remain intact. However, our numerical
results uncover remarkable differences between $\gamma>0$ and
$\gamma<0$ even when its value is small. In the extreme
circumstance when $C_{12}|\beta|\gg1$ and $C_{12}|\gamma|\gg1$, the
ground state is fully determined by the respective ratio of $\beta$
to $\gamma$, resulting in the same phase diagram as for
$C_1\beta_1=C_2\beta_2=0$, which is illustrated in Fig. \ref{fig2}.
The resulting phases are categorized as before and illustrated in Fig.
\ref{fig3}.

First, for two ferromagnetic condensates,
corresponding to Fig. \ref{fig1}(a) of $C_{12}\gamma=0$,
we find two phases: AA and FF
for $C_{12}\beta>0$ and $C_{12}\beta<0$ respectively.
The phase transition between them are discontinuous.
Increasing $C_{12}\gamma$ along the negative axis direction,
we find the same two phases AA and FF
if $C_{12}|\gamma|$ remains small in value, although the
phase boundary is found to shift with changing $C_{12}|\gamma|$.
Further increasing the strength of
singlet pairing interaction, a new phase MM emerges
between the AA and FF phases, where all four order parameters
evolve continuously to border the AA and FF phases.
Beyond a critical value of $C_{12}|\gamma|$,
the PP phase occupies
an interval section of the parameter $C_{12}\beta$.
Next, we increase $C_{12}\gamma$ from 0 along the positive
axis direction.
In the beginning, a positive $\gamma$-term decreases $|s_-|$.
No matter how small $C_{12}|\gamma|$ is,
the critical point (at $\gamma=0$) between the phases AA and FF
 diffuses into an interval,
and all four order parameters change continuously
from the FF to the AA phase.
When $C_{12}\gamma$ is sufficiently large, as in the case
of the opposite direction when $C_{12}\gamma<0$,
a special type of CC phase, which we call
CC1 ($\mathbf{f}_1^2=0$, $\mathbf{f}_2^2=1$)
arises over an interval of the parameter $C_{12}\beta$.
A continuous transition region is found to surround the CC1 phase
and borders the FF and AA phases. These results
are summarized in Fig. \ref{fig3}(a), with the
boundaries of continuous (discontinuous) phase transitions denoted by
blue (red) solid lines.

The case of two polar condensates
($0<C_1\beta_1<C_2\beta_2$) are shown in Fig. \ref{fig3}(b).
The $C_{12}\gamma=0$ line is seen to be partitioned into five intervals.
The middle interval around $C_{12}\beta=0$ is the region
where the two spins are completely independent as discussed before.
Additionally, two continuous changing intervals belong to the MM phase.
The remaining two intervals are the AA and FF phases,
the same as in the first case of Fig. \ref{fig3}(a).
The inclusion of the fourth competing interaction $\gamma$-term,
we find a new phase emerges on each side of the
$C_{12}\gamma=0$ line irrespective of the value of the
$\gamma$-term. This confirms the middle interval is
really a boundary between the PP phase and
a second special type of CC phase, denoted as
CC2 ($\mathbf{f}_1^2=0$, $\mathbf{f}_2^2=0$).
The transition across is found to be discontinuous.
When $C_{12}\gamma<0$, the PP phase is found to dominate.
In the opposite case, the CC2 phase wins.

The third case corresponds to a mixture of one ferromagnetic
($C_1\beta_1<0$) and one polar ($C_2\beta_2>0$) condensate.
When $C_{12}\gamma=0$ as discussed before,
there exist two phases AA and FF in both sides of
the $C_{12}\beta$ axis, plus a middle continuous
transition region. When $C_{12}\gamma\neq 0$,
we first decrease $C_{12}\gamma$ along the negative axis from 0.
Before reaching a critical value,
the ground state phase is found to be
essentially the same as for $C_{12}\gamma=0$.
Beyond this critical value, a PP phase emerges as
in the first case of Fig. \ref{fig3}(a).
Increasing $C_{12}\gamma\neq 0$ along the positive axis,
the results is found to change completely.
The critical value disappears.
A third type of CC phase, labeled as CC3
($\mathbf{f}_1^2=1$, $\mathbf{f}_2^2=0$),
emerges in the middle region. The transition between
the CC3 and the middle transition region MM phase
is discontinuous on both boundaries along the axis of $C_{12}\beta$,
denoted as green solid lines in Fig. \ref{fig3}(c).
This discontinuous phase transition boundary
is distinct from that of the red solid lines,
which denote boundaries between two phases with fixed order
parameters and the changes for all four order parameters
across the boundaries are exactly known.


Finally we hope to point out that although inter-species interaction
parameters are fixed, the interaction coefficients in the model
we consider such as $C_{12}\beta$ and $C_{12}\gamma$ remain tunable by controlling
optical trapping potentials or adjusting atom numbers.
In this study, we consider the special case of two equally populated
condensates inside fixed traps. In reality the effect of different atom numbers
can be used to adjust the four interaction coefficients.
We can also use different optical traps for the two species of atoms,
which can further enhance the tunability of the different ranges and ratios
of the values for $C_1$, $C_2$, and $C_{12}$.
\begin{figure}[H]
\centering
\includegraphics[width=3.0in]{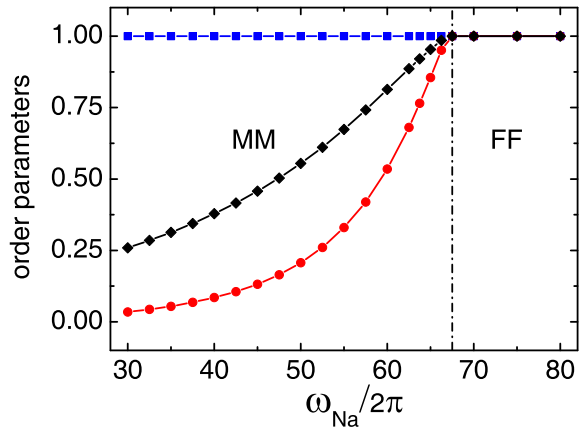}
\caption{(Color online). The dependence of the ground state order parameters on
the trap frequency $\omega_{\rm Na}$, at a fixed $\omega_{\rm Rb}=2\pi\times 50$Hz.
Blue squares, red circles,
and black diamonds denote respectively for the three order parameter $\mathbf{f}_1^2$,
$\mathbf{f}_2^2$, and $\mathbf{f}_1\cdot\mathbf{f}_2$. Lines are guides for the
eye obtained from connecting neighbouring data points.
Dot-dashed line specifies the phase boundary between
the MM and the FF phase.
} \label{fig4}
\end{figure}

As a concrete example, we consider here a realizable experiment to demonstrate
a quantum phase transition from the MM phase to the FF phase in a binary
mixture of $^{87}$Rb (species one) and $^{23}$Na (species two) spin-1 condensates.
The tuning is achieved through the control of the trapping frequency for
Na atoms. We assume $5\times 10^5$ $^{87}$Rb atoms and $10^5$ $^{23}$Na atoms
are confined harmonically $V_a(\vec{r})=\frac{1}{2}M_a\omega_ar^2$ with
$\omega_{\rm Rb}=2\pi\times 50$Hz, and $\omega_{\rm Na}$ tunable from
$2\pi\times 30$ to $2\pi\times 80$Hz. The inter-species atomic parameters
between $^{87}$Rb and $^{23}$Na are unknown. We therefore resort to the simple
approach of degenerate internal states \cite{pashov05,weiss03,stoof88} as the low energy
atomic interactions can be mostly attributed to the ground state configurations
of the two valence electrons. The inter-species scattering lengths for
singlet and triplet electronic states are approximately determined
already, given by $a_S=109 a_0$ and $a_T=70 a_0$
\cite{pashov05,weiss03}, where $a_0$ is the Bohr radius.
The inter-species interactions between
$^{87}$Rb and $^{23}$Na atoms are then parameterized by three
scattering lengths, each being a linear combination of
$a_S$ and $a_T$ weighted by the appropriate $9j$ coefficients
for the total combined spins of $F_{\rm tot}=0,1,$ and $2$ \cite{luo07}.
Within this approximation,
it's found coincidently \cite{luo07} that $\gamma$,
the parameter for inter-species singlet pairing
interaction is equal to zero.
The ground states are found through imaginary-time propagation
of the corresponding coupled Gross-Pitaevskii equations.
To compare with our former results within SMA, we redefine
the three order parameters as
$\mathbf{f}_1^2=\int d\vec{r} (\psi_i^*\vec{F}_{ij}\psi_j)^2/\int
d\vec{r} (\psi_i^*\psi_i)^2$, $\mathbf{f}_2^2=\int d\vec{r}
(\phi_i^*\vec{F}_{ij}\phi_j)^2/\int d\vec{r} (\phi_i^*\phi_i)^2$,
and $\mathbf{f}_1\cdot \mathbf{f}_2=\int d\vec{r}
(\psi_i^*\vec{F}_{ij}\psi_j) \cdot (\phi_i^*\vec{F}_{ij}\phi_j)/\int
d\vec{r} (\psi_i^*\psi_i)(\phi_i^*\phi_i)$, where $\psi_i$ and
$\phi_i$ ($i=1,0,-1$) are component wave functions respectively for
$^{87}$Rb and $^{23}$Na condensates. When the SMA is
satisfied, they reduce to that as defined in
Eq. (\ref{siener}), and the values of the order parameters
become the same as in Fig. \ref{fig1}(c).
This is indeed the case for most of our numerical calculations,
especially in the FF phase,
with all atoms fully spin polarized, or when two condensates
are uncorrelated.

In Fig. \ref{fig4}, the three order parameters for the ground state
are demonstrated under the changing trapping frequency of $^{23}$Na atoms.
With increasing value of $\omega_{\rm Na}$,
the ferromagnetic spin-spin coupling between $^{87}$Rb
and $^{23}$Na atoms will induce the spin polarization of $^{23}$Na atoms,
from initially unpolarized to a full polarization.
Compared to the phase diagram of a mixture with one ferromagnetic and one polar
spin-1 condensate as in Fig. \ref{fig3}(c), it's clear that by tuning
the trap frequency for the $^{23}$Na condensate, we induce a
quantum phase transition from the MM phase to the FF phase
in the mixture.

To conclude, we study the ground state of a binary mixture of two spin-1
condensates in the absence of B-field, using both analytical approaches
and the numerical method of simulated annealing.
Neglecting the singlet pairing interaction between the two species of atoms
($\gamma=0$), we present analytical formulae for all order parameters of the
ground state phase diagram including all phase boundaries and the
development of order parameters.
When $\gamma\neq0$, we find the phase diagram for the ground state at
three groups of specific values for the intra-species spin coupling
interactions. We suggest it is possible to tune to different phases,
or induce quantum phase transitions by changing optical traps
and/or atom numbers in the two condensates respectively.
Finally, we confirm that the mean field theory
we adopt remains an excellent approximation as all our results
have been repeated numerically with the full quantum approach of exact
diagonalization at small atom numbers.

This work is supported by US NSF, NSF of China under
Grant 10640420151 and 10774095, and NKBRSF of China under Grants
2006CB921206, 2006AA06Z104 and 2006CB921102.

\end{document}